\documentclass[a4,12pt,eclepsf]{article}
\usepackage{graphicx}
\usepackage{wrapfig, fancybox, fancyhdr, ascmac, amsmath, amssymb, makeidx, 
comment, multirow, bm}
\newcommand{\BM}{\begin{minipage}}
\newcommand{\EM}{\end{minipage}}

\begin{document}

\renewcommand{\theequation}
{\thesection.\arabic{equation}}
\thispagestyle{empty}
\vspace*{20mm} 
\begin{center}
{\LARGE {\bf The spectrum of low spin mesons at }} \\[5mm]
{\LARGE {\bf finite temperature in holographic}} \\[5mm]
{\LARGE {\bf noncommutative QCD}} \\

\vspace*{20mm}

\renewcommand{\thefootnote}{\fnsymbol{footnote}}

{\Large Tadahito NAKAJIMA, \footnotemark[1]\,
Yukiko OHTAKE \footnotemark[2]
and \,Kenji SUZUKI \footnotemark[3]} \\
\vspace*{20mm}

$\footnotemark[1]$ {\it College of Engineering, Nihon University, 
Fukushima 963-8642, Japan} \\[2mm]
$\footnotemark[2]$
{\it Toyama National College of Technology, Toyama 939-8630, Japan} \\[2mm]
$\footnotemark[3]$
% Faculty of Liberal Arts \& Science, 
{\it Department of Physics, Ochanomizu University, 
Tokyo 112-8610, Japan} \\[15mm]

%{\Large \today} \\[15mm]

{\bf Abstract} \\[10mm]

\end{center}

We have constructed a noncommutative deformation of the holographic QCD (Sakai-Sugimoto) model and evaluated the mass spectrum of low spin vector mesons at 
finite temperature. The masses of light vector- and pseudovector-meson in the 
noncommutative holographic QCD model reduces over the whole area in the 
intermediate-temperature regime compared to the commutative case. However, the 
space noncommutativity does not change the properties of temperature 
dependence for the mass spectrum of low spin mesons. The masses of meson also 
decrease with increasing temperature in noncommutative case. 

\clearpage

\setcounter{section}{0}
\section{Introduction}
\setcounter{page}{1}
\setcounter{equation}{0}

Noncommutative gauge theories (gauge theories on noncommutative space) 
naturally arise as low energy theories of D-branes in 
Neveu-Schwarz-Neveu-Schwarz (NS-NS) $B$-field background 
\cite{CDS, DH, AASJ, SJ, SW}. The space noncommutativity brings nontrivial 
properties on the gauge field theory at the quantum level. A remarkable 
example is the mixing between the infrared and the ultraviolet degrees of 
freedom. Although the product of the momentum and 
the noncommutativity parameter plays the role of ultraviolet cut-off, the 
result is singular when the momentum or the noncommutativity parameter is taken to zero. This phenomenon is referred to as the UV/IR mixing \cite{MRS}. Although the noncommutative gauge theories have been studied extensively, it is hard to investigate them in the perturbative approach. Little is currently known of the 
non-perturbative properties of noncommutative gauge theories.

The noncommutative gauge theories have gravity duals whose near horizon region 
describes the large $N$ limit of the noncommutative gauge theories 
\cite{HI, MR, AOSJ}. Based on the noncommutative deformation of gauge/gravity 
duality, we can investigate the non-perturbative aspects of the large $N$ 
limit of the noncommutative gauge theories. For instance, the space 
noncommutativity modifies the Wilson loop behavior \cite{dhar_kita, lee_sin, 
taka_naka-suzu} and glueball mass spectra \cite{NST}. The gravity duals of 
noncommutative gauge theories with matter in the fundamental representation 
have also been constructed by adding probe flavor branes \cite{APR}. 
Employing the gravity dual description of noncommutative gauge theories 
with flavor degrees of freedom we have been able to find the space 
noncommutativity is also reflected in the flavor dynamics. For instance, the 
mass spectrum of mesons can be modified by the space noncommutativity \cite{APR}.

Fundamental properties of quantum chromodynamics (QCD) at low energies are 
confinement and chiral symmetry breaking. The holographic QCD model (D4-D8-
$\overline{\rm D8}$ model) so-called Sakai--Sugimoto model has been known to 
capture these properties of QCD at low energies \cite{SS1, SS2}. The 
holographic QCD models can be modified to introduce a finite temperature. The 
phase transition of chiral symmetry restoration could be described in terms of 
the holographic QCD at finite temperature. The holographic QCD at finite 
temperature has three phase structures, confined with broken chiral symmetry, 
deconfined with restored chiral symmetry, and deconfined with broken chiral 
symmetry \cite{ASY, PS}. The finite baryon chemical potential and density can 
be also naturally introduced in the holographic QCD at finite temperature and 
bring the rich structure in this model \cite{HT, PS2, Y, OB_GL_ML, 
MR_HHS_MVR_JW}. The properties of mesons at finite temperature has been analyzed within the framework of the holographic QCD at finite temperature 
\cite{PSZ, PSZ2, KSZ}. How does the influence of the space noncommutativity 
reflect the properties of the mesons in the QCD at finite temperature.

In this paper, we construct the noncommutative deformation of the holographic 
QCD at finite temperature in imitation of the method by Alishahiha et al. \cite{AOSJ} and Arean et al. \cite{APR} and investigate the influence of space 
noncommutativity on the mass spectrum of low spin mesons in the QCD at finite 
temperature. There are some attempts to investigate the nontrivial effect of 
the space noncommutativity on the QCD in the context of the noncommutative 
deformation of the holographic QCD \cite{TN_YO_KS, M_AA}. As will be seen 
later, the space noncommutativity is also able to modify the mass spectrum of 
meson.

This paper is organized as follows. In section 2, we construct the 
noncommutative deformation of the holographic QCD model at finite temperature. In section 3, we analyze the mass spectrum of low-spin meson in the intermediate-temperature phase, which is the phase of deconfined with broken chiral 
symmetry within the framework of the noncommutative deformation of the holographic QCD model. Section 4 is devoted to conclusions. 

%
%%%%%%%%%%%%%%%%%%%%%%%%%%%%%%%%%%%%%%%%%%%%%%%%%%%%%%%%%%%%%%%%%%%%%%%%%%%%%%%
%

\section{Noncommutative deformation of the holographic QCD model at finite 
temperature} 

\setcounter{equation}{0}
\addtocounter{enumi}{1}

In this section, we consider a noncommutative deformation of the holographic 
QCD (Sakai--Sugimoto) model at finite temperature based on the prescription 
of Arean--Paredes--Ramallo \cite{APR}. 
%Maldacena and Russo \cite{MR}. 
The holographic QCD model is a gravity dual for a $4+1$ dimensional QCD with 
${\rm U}({\rm N}_{f})_{L} \times {\rm U}({\rm N}_{f})_{R}$ global chiral 
symmetry whose symmetry is spontaneously broken \cite{SS1, SS2}. This model is 
a $\text{D4-D8-}\overline{\text{D8}}\text{-}$brane system consisting $S^{1}$ 
compactified $N_{c}$ D4-branes and $N_{f}$ $\text{D8-}\overline{\text{D8}}$ 
pairs transverse to the $S^{1}$. The near-horizon limit of the set of $N_{c}$ 
D4-branes solution compactified on $S^{1}$ takes the form
\begin{align}
ds^{2} &=\left(\frac{U}{R_{\rm D4}}\right)^{3/2}\Bigl(-(dt)^{2}+(dx^{1})^{2}
+ (dx^{2})^{2}+(dx^{3})^{2} + f_{\rm KK}(U)\,d\tau^{2} \Bigr) \nonumber \\
& + \left(\frac{R_{\rm D4}}{U}\right)^{3/2} 
\left(\frac{dU^{2}}{f_{\rm KK}(U)}+U^{2}d\Omega_{4}^{2} \right)\,,\nonumber \\
& R_{\rm D4}^{3}=\pi g_{s}N_{c}l^{3}_{s}\,, \qquad 
f_{\rm KK}(U)=1-\frac{U_{\rm KK}{}^{3}}{U^{3}}\,,
\label{201}
\end{align}
where $U_{\rm KK}$ is a parameter, $U$ is the radial direction bounded from below by $U \geq U_{\rm KK}$, $\tau$ is compactified direction of the D4-brane world volume which is transverse to the $\text{D8-}\overline{\text{D8}}$-branes, $g_{s}$ and $l_{s}$ are the string coupling and length respectively. The dilaton $\phi$ and the field strength $F_{4}$ of the RR 3-form $C_{3}$ are given by 
\begin{align}
e^{\phi}=g_{s}\left(\dfrac{U}{R_{\rm D4}}\right)^{3/4}\,, \qquad 
F_{4}=dC_{3}=\dfrac{2\pi N_{c}}{V_{4}}\epsilon_{4}\,,
\label{202}
\end{align}
where $V_{4}=8\pi^{2}/3$ is the volume of unit $S^{4}$ and $\epsilon_{4}$ is the corresponding volume form. In order to avoid a conical singularity at $U=U_{\rm KK}$, the $\tau$ direction should have a period of 
%The tip of the cigar is non-singular if and only if the periodicity of $\tau$ 
%is
%
\begin{align}
\delta \tau 
= \dfrac{4\pi}{3}\left(\dfrac{R_{\rm D4}^{3}}{U_{\rm KK}}\right)^{1/2}
= 2\pi R = \dfrac{2\pi}{M_{\rm KK}}\,,
\label{203}
\end{align}
where $R$ is radius of $S^{1}$ and $M_{\rm KK}$ is the Kaluza--Klein mass. The parameter $U_{\rm KK}$ is related to the Kaluza--Klein mass $M_{\rm KK}$ via the relation (\ref{203}). The five dimensional gauge coupling is expressed in terms of $g_{s}$ and $l_{s}$ as $g_{\rm YM}^{2}=(2\pi)^{2}g_{s}l_{s}$. The gravity description is valid for strong coupling $\lambda \gg R$, where as usual $\lambda = g_{\rm YM}^{2}N_{c}$ denotes the 't Hooft coupling. 

The holographic QCD model at finite temperature has been proposed in \cite{ASY, PS, PSZ}. In order to introduce a finite temperature $T$ in the model, we consider the Euclidean gravitational solution which is asymptotically equals to (\ref{201}) but with the compactification of Euclidean time direction $t_{E}$. In this solution the periodicity of $t_{E}$ is arbitrary and equals to $\beta=1/T$.

Another solution with the same asymptotic is given by interchanging the role of $t_{E}$ and $\tau$ directions, 
\begin{align}
ds^{2} &=\left(\frac{U}{R_{\rm D4}}\right)^{3/2}
\Bigl(f_{\rm T}(U)\,(dt_{E})^{2}+(dx^{1})^{2}
+ (dx^{2})^{2}+(dx^{3})^{2} + d\tau^{2} \Bigr) \nonumber \\
& + \left(\frac{R_{\rm D4}}{U}\right)^{3/2}
\left(\frac{dU^{2}}{f_{\rm T}(U)}+U^{2}d\Omega_{4}^{2} \right)\,,\nonumber \\
& R_{\rm D4}^{3}=\pi g_{s}N_{c}l^{3}_{s}\,, \qquad 
f_{\rm T}(U)=1-\frac{U^{3}_{T}}{U^{3}}\,,
\label{204}
\end{align}
where $U_{T}$ is a parameter. To avoid a singularity at $U=U_{T}$ the period of $\delta t_{E}$ of the compactified time direction is set to 
\begin{align}
\delta t_{E} = \dfrac{4\pi}{3}\left(\dfrac{R_{\rm D4}^{3}}{U_{T}}\right)^{1/2}
=\dfrac{1}{T}\,,
\label{205}
\end{align}
and the parameter $U_{T}$ is related to the temperature $T$. The metric 
(\ref{201}) with the compactification of Euclidean time $t_{E}$ is dominant in 
the low temperature $T <1/2\pi R$, while the metric (\ref{204}) is dominant in 
the high temperature $T >1/2\pi R$. The transition between the metric 
(\ref{201}) and the metric (\ref{204}) happens when $T=T_{c}=1/2\pi R$. This 
transition is first-order and corresponds to the confinement/deconfinement 
phase transition in the gauge theory side. \\

\begin{center}
\scalebox{0.6}{%WinTpicVersion3.08
\unitlength 0.1in
\begin{picture}( 53.0500, 24.1500)(  1.9500,-30.3000)
% ELLIPSE 2 0 3 0
% 4 4300 1245 3500 1045 3500 1045 3500 1045
% 
\special{pn 8}%
\special{ar 4300 1246 800 200  0.0000000 6.2831853}%
% STR 2 0 3 0
% 3 4300 745 4300 845 5 0
% \LARGE $x_{0}$
\put(43.0000,-8.4500){\makebox(0,0){\LARGE $x_{0}$}}%
% ELLIPSE 2 0 3 0
% 4 4300 2645 5100 2845 2730 2645 5330 2645
% 
\special{pn 8}%
\special{ar 4300 2646 800 200  6.2831853 6.2831853}%
\special{ar 4300 2646 800 200  0.0000000 3.1415927}%
% LINE 2 0 3 0
% 4 3500 1245 3500 2645 5100 1245 5100 2645
% 
\special{pn 8}%
\special{pa 3500 1246}%
\special{pa 3500 2646}%
\special{fp}%
\special{pa 5100 1246}%
\special{pa 5100 2646}%
\special{fp}%
% STR 2 0 3 0
% 3 1300 600 1300 700 5 0
% \LARGE $U$
\put(13.0000,-7.0000){\makebox(0,0){\LARGE $U$}}%
% ELLIPSE 2 0 3 0
% 4 2300 1250 1500 1050 1500 1050 1500 1050
% 
\special{pn 8}%
\special{ar 2300 1250 800 200  0.0000000 6.2831853}%
% ELLIPSE 2 0 3 0
% 4 2300 1250 3100 2850 900 1250 3100 1250
% 
\special{pn 8}%
\special{ar 2300 1250 800 1600  6.2831853 6.2831853}%
\special{ar 2300 1250 800 1600  0.0000000 3.1415927}%
% ELLIPSE 1 0 3 0
% 4 2300 1390 2900 2590 1250 1390 2900 1390
% 
\special{pn 13}%
\special{ar 2300 1390 600 1200  6.2831853 6.2831853}%
\special{ar 2300 1390 600 1200  0.0000000 3.1415927}%
% STR 2 0 3 0
% 3 1050 2750 1050 2850 5 0
% \LARGE $U_{\rm KK}$
\put(10.5000,-28.5000){\makebox(0,0){\LARGE $U_{\rm KK}$}}%
% STR 2 0 3 0
% 3 1100 2500 1100 2600 5 0
% \LARGE $U_{0}$
\put(11.0000,-26.0000){\makebox(0,0){\LARGE $U_{0}$}}%
% STR 2 0 3 0
% 3 2300 750 2300 850 5 0
% \LARGE $x_{4}$
\put(23.0000,-8.5000){\makebox(0,0){\LARGE $x_{4}$}}%
% STR 2 0 3 0
% 3 2300 2100 2300 2200 5 0
% \LARGE ${\rm D8}\text{-}\overline{\rm D8}$
\put(23.0000,-22.0000){\makebox(0,0){\LARGE ${\rm D8}\text{-}\overline{\rm D8}$}}%
% VECTOR 2 0 3 0
% 2 1300 3030 1300 830
% 
\special{pn 8}%
\special{pa 1300 3030}%
\special{pa 1300 830}%
\special{fp}%
\special{sh 1}%
\special{pa 1300 830}%
\special{pa 1280 898}%
\special{pa 1300 884}%
\special{pa 1320 898}%
\special{pa 1300 830}%
\special{fp}%
% LINE 2 2 3 0
% 2 1300 2600 5500 2600
% 
\special{pn 8}%
\special{pa 1300 2600}%
\special{pa 5500 2600}%
\special{dt 0.045}%
% ELLIPSE 2 0 3 0
% 4 4300 2400 5100 2600 3500 2400 5100 2400
% 
\special{pn 8}%
\special{ar 4300 2400 800 200  6.2831853 6.2831853}%
\special{ar 4300 2400 800 200  0.0000000 3.1415927}%
% LINE 1 2 3 0
% 60 4970 2510 3870 1410 4920 2520 3800 1400 4880 2540 3720 1380 4830 2550 3630 1350 4780 2560 3500 1280 4730 2570 3500 1340 4670 2570 3500 1400 4620 2580 3500 1460 4570 2590 3500 1520 4510 2590 3500 1580 4450 2590 3500 1640 4390 2590 3500 1700 4330 2590 3500 1760 4270 2590 3500 1820 4210 2590 3500 1880 4150 2590 3500 1940 4090 2590 3500 2000 4030 2590 3500 2060 3960 2580 3500 2120 3890 2570 3500 2180 3820 2560 3500 2240 3740 2540 3500 2300 3660 2520 3500 2360 5010 2490 3940 1420 5050 2470 4010 1430 5080 2440 4070 1430 5100 2400 4140 1440 5100 2340 4200 1440 5100 2280 4260 1440 5100 2220 4320 1440
% 
\special{pn 13}%
\special{pa 4970 2510}%
\special{pa 3870 1410}%
\special{dt 0.045}%
\special{pa 4920 2520}%
\special{pa 3800 1400}%
\special{dt 0.045}%
\special{pa 4880 2540}%
\special{pa 3720 1380}%
\special{dt 0.045}%
\special{pa 4830 2550}%
\special{pa 3630 1350}%
\special{dt 0.045}%
\special{pa 4780 2560}%
\special{pa 3500 1280}%
\special{dt 0.045}%
\special{pa 4730 2570}%
\special{pa 3500 1340}%
\special{dt 0.045}%
\special{pa 4670 2570}%
\special{pa 3500 1400}%
\special{dt 0.045}%
\special{pa 4620 2580}%
\special{pa 3500 1460}%
\special{dt 0.045}%
\special{pa 4570 2590}%
\special{pa 3500 1520}%
\special{dt 0.045}%
\special{pa 4510 2590}%
\special{pa 3500 1580}%
\special{dt 0.045}%
\special{pa 4450 2590}%
\special{pa 3500 1640}%
\special{dt 0.045}%
\special{pa 4390 2590}%
\special{pa 3500 1700}%
\special{dt 0.045}%
\special{pa 4330 2590}%
\special{pa 3500 1760}%
\special{dt 0.045}%
\special{pa 4270 2590}%
\special{pa 3500 1820}%
\special{dt 0.045}%
\special{pa 4210 2590}%
\special{pa 3500 1880}%
\special{dt 0.045}%
\special{pa 4150 2590}%
\special{pa 3500 1940}%
\special{dt 0.045}%
\special{pa 4090 2590}%
\special{pa 3500 2000}%
\special{dt 0.045}%
\special{pa 4030 2590}%
\special{pa 3500 2060}%
\special{dt 0.045}%
\special{pa 3960 2580}%
\special{pa 3500 2120}%
\special{dt 0.045}%
\special{pa 3890 2570}%
\special{pa 3500 2180}%
\special{dt 0.045}%
\special{pa 3820 2560}%
\special{pa 3500 2240}%
\special{dt 0.045}%
\special{pa 3740 2540}%
\special{pa 3500 2300}%
\special{dt 0.045}%
\special{pa 3660 2520}%
\special{pa 3500 2360}%
\special{dt 0.045}%
\special{pa 5010 2490}%
\special{pa 3940 1420}%
\special{dt 0.045}%
\special{pa 5050 2470}%
\special{pa 4010 1430}%
\special{dt 0.045}%
\special{pa 5080 2440}%
\special{pa 4070 1430}%
\special{dt 0.045}%
\special{pa 5100 2400}%
\special{pa 4140 1440}%
\special{dt 0.045}%
\special{pa 5100 2340}%
\special{pa 4200 1440}%
\special{dt 0.045}%
\special{pa 5100 2280}%
\special{pa 4260 1440}%
\special{dt 0.045}%
\special{pa 5100 2220}%
\special{pa 4320 1440}%
\special{dt 0.045}%
% LINE 1 2 3 1
% 30 5100 2160 4380 1440 5100 2100 4440 1440 5100 2040 4490 1430 5100 1980 4550 1430 5100 1920 4610 1430 5100 1860 4660 1420 5100 1800 4710 1410 5100 1740 4760 1400 5100 1680 4810 1390 5100 1620 4860 1380 5100 1560 4910 1370 5100 1500 4960 1360 5100 1440 5000 1340 5100 1380 5040 1320 5100 1320 5070 1290
% 
\special{pn 13}%
\special{pa 5100 2160}%
\special{pa 4380 1440}%
\special{dt 0.045}%
\special{pa 5100 2100}%
\special{pa 4440 1440}%
\special{dt 0.045}%
\special{pa 5100 2040}%
\special{pa 4490 1430}%
\special{dt 0.045}%
\special{pa 5100 1980}%
\special{pa 4550 1430}%
\special{dt 0.045}%
\special{pa 5100 1920}%
\special{pa 4610 1430}%
\special{dt 0.045}%
\special{pa 5100 1860}%
\special{pa 4660 1420}%
\special{dt 0.045}%
\special{pa 5100 1800}%
\special{pa 4710 1410}%
\special{dt 0.045}%
\special{pa 5100 1740}%
\special{pa 4760 1400}%
\special{dt 0.045}%
\special{pa 5100 1680}%
\special{pa 4810 1390}%
\special{dt 0.045}%
\special{pa 5100 1620}%
\special{pa 4860 1380}%
\special{dt 0.045}%
\special{pa 5100 1560}%
\special{pa 4910 1370}%
\special{dt 0.045}%
\special{pa 5100 1500}%
\special{pa 4960 1360}%
\special{dt 0.045}%
\special{pa 5100 1440}%
\special{pa 5000 1340}%
\special{dt 0.045}%
\special{pa 5100 1380}%
\special{pa 5040 1320}%
\special{dt 0.045}%
\special{pa 5100 1320}%
\special{pa 5070 1290}%
\special{dt 0.045}%
% ELLIPSE 2 0 3 0
% 4 2300 1450 3100 1650 1500 1650 3100 1650
% 
\special{pn 8}%
\special{ar 2300 1450 800 200  0.7853982 2.3561945}%
% SARROW 2 0 3 1
% 2 1752 1596 1734 1591
% 
\special{pn 8}%
\special{pa 1752 1596}%
\special{pa 1734 1592}%
\special{fp}%
\special{sh 1}%
\special{pa 1734 1592}%
\special{pa 1794 1628}%
\special{pa 1786 1606}%
\special{pa 1804 1590}%
\special{pa 1734 1592}%
\special{fp}%
% SARROW 2 0 3 2
% 2 2848 1596 2866 1591
% 
\special{pn 8}%
\special{pa 2848 1596}%
\special{pa 2866 1592}%
\special{fp}%
\special{sh 1}%
\special{pa 2866 1592}%
\special{pa 2796 1590}%
\special{pa 2816 1606}%
\special{pa 2808 1628}%
\special{pa 2866 1592}%
\special{fp}%
% STR 2 0 3 0
% 3 2300 1700 2300 1800 5 0
% \LARGE $L$
\put(23.0000,-18.0000){\makebox(0,0){\LARGE $L$}}%
% VECTOR 2 0 3 0
% 4 2900 1250 3100 1250 2500 1250 2300 1250
% 
\special{pn 8}%
\special{pa 2900 1250}%
\special{pa 3100 1250}%
\special{fp}%
\special{sh 1}%
\special{pa 3100 1250}%
\special{pa 3034 1230}%
\special{pa 3048 1250}%
\special{pa 3034 1270}%
\special{pa 3100 1250}%
\special{fp}%
\special{pa 2500 1250}%
\special{pa 2300 1250}%
\special{fp}%
\special{sh 1}%
\special{pa 2300 1250}%
\special{pa 2368 1270}%
\special{pa 2354 1250}%
\special{pa 2368 1230}%
\special{pa 2300 1250}%
\special{fp}%
% STR 2 0 3 0
% 3 2700 1150 2700 1250 5 0
% \LARGE $R$
\put(27.0000,-12.5000){\makebox(0,0){\LARGE $R$}}%
% VECTOR 2 0 3 0
% 4 4900 1250 5100 1250 4500 1250 4300 1250
% 
\special{pn 8}%
\special{pa 4900 1250}%
\special{pa 5100 1250}%
\special{fp}%
\special{sh 1}%
\special{pa 5100 1250}%
\special{pa 5034 1230}%
\special{pa 5048 1250}%
\special{pa 5034 1270}%
\special{pa 5100 1250}%
\special{fp}%
\special{pa 4500 1250}%
\special{pa 4300 1250}%
\special{fp}%
\special{sh 1}%
\special{pa 4300 1250}%
\special{pa 4368 1270}%
\special{pa 4354 1250}%
\special{pa 4368 1230}%
\special{pa 4300 1250}%
\special{fp}%
% STR 2 0 3 0
% 3 4700 1150 4700 1250 5 0
% \LARGE $R'$
\put(47.0000,-12.5000){\makebox(0,0){\LARGE $R'$}}%
% LINE 2 2 3 0
% 2 1300 2850 5500 2850
% 
\special{pn 8}%
\special{pa 1300 2850}%
\special{pa 5500 2850}%
\special{dt 0.045}%
\end{picture}%
} \\
{\bf Fig. 1}\; The $\text{D8-}\overline{\text{D8}}$-branes configurations at low temperature. \\[10mm]
\end{center}

\begin{center}
\begin{tabular}{cc}
\scalebox{0.6}{%WinTpicVersion3.08
\unitlength 0.1in
\begin{picture}( 52.3000, 24.1500)(  2.7000,-30.3000)
% ELLIPSE 2 0 3 0
% 4 4300 1245 3500 1045 3500 1045 3500 1045
% 
\special{pn 8}%
\special{ar 4300 1246 800 200  0.0000000 6.2831853}%
% STR 2 0 3 0
% 3 4300 745 4300 845 5 0
% \LARGE $x_{0}$
\put(43.0000,-8.4500){\makebox(0,0){\LARGE $x_{0}$}}%
% STR 2 0 3 0
% 3 1300 600 1300 700 5 0
% \LARGE $U$
\put(13.0000,-7.0000){\makebox(0,0){\LARGE $U$}}%
% ELLIPSE 2 0 3 0
% 4 2300 1250 1500 1050 1500 1050 1500 1050
% 
\special{pn 8}%
\special{ar 2300 1250 800 200  0.0000000 6.2831853}%
% ELLIPSE 2 0 3 0
% 4 4300 1250 5100 2850 2900 1250 5100 1250
% 
\special{pn 8}%
\special{ar 4300 1250 800 1600  6.2831853 6.2831853}%
\special{ar 4300 1250 800 1600  0.0000000 3.1415927}%
% STR 2 0 3 0
% 3 1080 2750 1080 2850 5 0
% \LARGE $U_{\rm T}$
\put(10.8000,-28.5000){\makebox(0,0){\LARGE $U_{\rm T}$}}%
% STR 2 0 3 0
% 3 2300 750 2300 850 5 0
% \LARGE $x_{4}$
\put(23.0000,-8.5000){\makebox(0,0){\LARGE $x_{4}$}}%
% STR 2 0 3 0
% 3 2300 2100 2300 2200 5 0
% \LARGE ${\rm D8}\text{-}\overline{\rm D8}$
\put(23.0000,-22.0000){\makebox(0,0){\LARGE ${\rm D8}\text{-}\overline{\rm D8}$}}%
% VECTOR 2 0 3 0
% 2 1300 3030 1300 830
% 
\special{pn 8}%
\special{pa 1300 3030}%
\special{pa 1300 830}%
\special{fp}%
\special{sh 1}%
\special{pa 1300 830}%
\special{pa 1280 898}%
\special{pa 1300 884}%
\special{pa 1320 898}%
\special{pa 1300 830}%
\special{fp}%
% LINE 2 2 3 0
% 2 1300 2850 5500 2850
% 
\special{pn 8}%
\special{pa 1300 2850}%
\special{pa 5500 2850}%
\special{dt 0.045}%
% ELLIPSE 2 0 3 0
% 4 2300 1450 3100 1650 1500 1650 3100 1650
% 
\special{pn 8}%
\special{ar 2300 1450 800 200  0.7853982 2.3561945}%
% SARROW 2 0 3 1
% 2 1752 1596 1734 1591
% 
\special{pn 8}%
\special{pa 1752 1596}%
\special{pa 1734 1592}%
\special{fp}%
\special{sh 1}%
\special{pa 1734 1592}%
\special{pa 1794 1628}%
\special{pa 1786 1606}%
\special{pa 1804 1590}%
\special{pa 1734 1592}%
\special{fp}%
% SARROW 2 0 3 2
% 2 2848 1596 2866 1591
% 
\special{pn 8}%
\special{pa 2848 1596}%
\special{pa 2866 1592}%
\special{fp}%
\special{sh 1}%
\special{pa 2866 1592}%
\special{pa 2796 1590}%
\special{pa 2816 1606}%
\special{pa 2808 1628}%
\special{pa 2866 1592}%
\special{fp}%
% STR 2 0 3 0
% 3 2300 1700 2300 1800 5 0
% \LARGE $L$
\put(23.0000,-18.0000){\makebox(0,0){\LARGE $L$}}%
% VECTOR 2 0 3 0
% 4 2900 1250 3100 1250 2500 1250 2300 1250
% 
\special{pn 8}%
\special{pa 2900 1250}%
\special{pa 3100 1250}%
\special{fp}%
\special{sh 1}%
\special{pa 3100 1250}%
\special{pa 3034 1230}%
\special{pa 3048 1250}%
\special{pa 3034 1270}%
\special{pa 3100 1250}%
\special{fp}%
\special{pa 2500 1250}%
\special{pa 2300 1250}%
\special{fp}%
\special{sh 1}%
\special{pa 2300 1250}%
\special{pa 2368 1270}%
\special{pa 2354 1250}%
\special{pa 2368 1230}%
\special{pa 2300 1250}%
\special{fp}%
% STR 2 0 3 0
% 3 2700 1150 2700 1250 5 0
% \LARGE $R$
\put(27.0000,-12.5000){\makebox(0,0){\LARGE $R$}}%
% VECTOR 2 0 3 0
% 4 4900 1250 5100 1250 4500 1250 4300 1250
% 
\special{pn 8}%
\special{pa 4900 1250}%
\special{pa 5100 1250}%
\special{fp}%
\special{sh 1}%
\special{pa 5100 1250}%
\special{pa 5034 1230}%
\special{pa 5048 1250}%
\special{pa 5034 1270}%
\special{pa 5100 1250}%
\special{fp}%
\special{pa 4500 1250}%
\special{pa 4300 1250}%
\special{fp}%
\special{sh 1}%
\special{pa 4300 1250}%
\special{pa 4368 1270}%
\special{pa 4354 1250}%
\special{pa 4368 1230}%
\special{pa 4300 1250}%
\special{fp}%
% STR 2 0 3 0
% 3 4700 1150 4700 1250 5 0
% \LARGE $R'$
\put(47.0000,-12.5000){\makebox(0,0){\LARGE $R'$}}%
% LINE 1 0 3 0
% 2 1720 1390 1720 2790
% 
\special{pn 13}%
\special{pa 1720 1390}%
\special{pa 1720 2790}%
\special{fp}%
% LINE 1 0 3 0
% 2 2900 1380 2900 2780
% 
\special{pn 13}%
\special{pa 2900 1380}%
\special{pa 2900 2780}%
\special{fp}%
% ELLIPSE 2 0 3 0
% 4 2300 2650 3100 2850 730 2650 3330 2650
% 
\special{pn 8}%
\special{ar 2300 2650 800 200  6.2831853 6.2831853}%
\special{ar 2300 2650 800 200  0.0000000 3.1415927}%
% LINE 2 0 3 0
% 4 1510 1250 1510 2650 3110 1250 3110 2650
% 
\special{pn 8}%
\special{pa 1510 1250}%
\special{pa 1510 2650}%
\special{fp}%
\special{pa 3110 1250}%
\special{pa 3110 2650}%
\special{fp}%
% LINE 1 2 3 0
% 60 4690 2650 3510 1470 4710 2610 3510 1410 4740 2580 3510 1350 4770 2550 3500 1280 4790 2510 3630 1350 4810 2470 3720 1380 4830 2430 3800 1400 4860 2400 3870 1410 4880 2360 3940 1420 4890 2310 4010 1430 4910 2270 4070 1430 4930 2230 4140 1440 4950 2190 4200 1440 4960 2140 4260 1440 4980 2100 4320 1440 4990 2050 4380 1440 5000 2000 4440 1440 5010 1950 4490 1430 5030 1910 4550 1430 5040 1860 4610 1430 5050 1810 4660 1420 5050 1750 4710 1410 5060 1700 4760 1400 5070 1650 4810 1390 5080 1600 4860 1380 5080 1540 4910 1370 5090 1490 4960 1360 5090 1430 5000 1340 5090 1370 5040 1320 5100 1320 5070 1290
% 
\special{pn 13}%
\special{pa 4690 2650}%
\special{pa 3510 1470}%
\special{dt 0.045}%
\special{pa 4710 2610}%
\special{pa 3510 1410}%
\special{dt 0.045}%
\special{pa 4740 2580}%
\special{pa 3510 1350}%
\special{dt 0.045}%
\special{pa 4770 2550}%
\special{pa 3500 1280}%
\special{dt 0.045}%
\special{pa 4790 2510}%
\special{pa 3630 1350}%
\special{dt 0.045}%
\special{pa 4810 2470}%
\special{pa 3720 1380}%
\special{dt 0.045}%
\special{pa 4830 2430}%
\special{pa 3800 1400}%
\special{dt 0.045}%
\special{pa 4860 2400}%
\special{pa 3870 1410}%
\special{dt 0.045}%
\special{pa 4880 2360}%
\special{pa 3940 1420}%
\special{dt 0.045}%
\special{pa 4890 2310}%
\special{pa 4010 1430}%
\special{dt 0.045}%
\special{pa 4910 2270}%
\special{pa 4070 1430}%
\special{dt 0.045}%
\special{pa 4930 2230}%
\special{pa 4140 1440}%
\special{dt 0.045}%
\special{pa 4950 2190}%
\special{pa 4200 1440}%
\special{dt 0.045}%
\special{pa 4960 2140}%
\special{pa 4260 1440}%
\special{dt 0.045}%
\special{pa 4980 2100}%
\special{pa 4320 1440}%
\special{dt 0.045}%
\special{pa 4990 2050}%
\special{pa 4380 1440}%
\special{dt 0.045}%
\special{pa 5000 2000}%
\special{pa 4440 1440}%
\special{dt 0.045}%
\special{pa 5010 1950}%
\special{pa 4490 1430}%
\special{dt 0.045}%
\special{pa 5030 1910}%
\special{pa 4550 1430}%
\special{dt 0.045}%
\special{pa 5040 1860}%
\special{pa 4610 1430}%
\special{dt 0.045}%
\special{pa 5050 1810}%
\special{pa 4660 1420}%
\special{dt 0.045}%
\special{pa 5050 1750}%
\special{pa 4710 1410}%
\special{dt 0.045}%
\special{pa 5060 1700}%
\special{pa 4760 1400}%
\special{dt 0.045}%
\special{pa 5070 1650}%
\special{pa 4810 1390}%
\special{dt 0.045}%
\special{pa 5080 1600}%
\special{pa 4860 1380}%
\special{dt 0.045}%
\special{pa 5080 1540}%
\special{pa 4910 1370}%
\special{dt 0.045}%
\special{pa 5090 1490}%
\special{pa 4960 1360}%
\special{dt 0.045}%
\special{pa 5090 1430}%
\special{pa 5000 1340}%
\special{dt 0.045}%
\special{pa 5090 1370}%
\special{pa 5040 1320}%
\special{dt 0.045}%
\special{pa 5100 1320}%
\special{pa 5070 1290}%
\special{dt 0.045}%
% LINE 1 2 3 1
% 24 4660 2680 3510 1530 4630 2710 3520 1600 4590 2730 3530 1670 4560 2760 3540 1740 4520 2780 3550 1810 4480 2800 3570 1890 4440 2820 3590 1970 4400 2840 3610 2050 4350 2850 3640 2140 4290 2850 3670 2230 4220 2840 3720 2340 4130 2810 3780 2460
% 
\special{pn 13}%
\special{pa 4660 2680}%
\special{pa 3510 1530}%
\special{dt 0.045}%
\special{pa 4630 2710}%
\special{pa 3520 1600}%
\special{dt 0.045}%
\special{pa 4590 2730}%
\special{pa 3530 1670}%
\special{dt 0.045}%
\special{pa 4560 2760}%
\special{pa 3540 1740}%
\special{dt 0.045}%
\special{pa 4520 2780}%
\special{pa 3550 1810}%
\special{dt 0.045}%
\special{pa 4480 2800}%
\special{pa 3570 1890}%
\special{dt 0.045}%
\special{pa 4440 2820}%
\special{pa 3590 1970}%
\special{dt 0.045}%
\special{pa 4400 2840}%
\special{pa 3610 2050}%
\special{dt 0.045}%
\special{pa 4350 2850}%
\special{pa 3640 2140}%
\special{dt 0.045}%
\special{pa 4290 2850}%
\special{pa 3670 2230}%
\special{dt 0.045}%
\special{pa 4220 2840}%
\special{pa 3720 2340}%
\special{dt 0.045}%
\special{pa 4130 2810}%
\special{pa 3780 2460}%
\special{dt 0.045}%
\end{picture}%
} & \scalebox{0.6}{%WinTpicVersion3.08
\unitlength 0.1in
\begin{picture}( 52.1000, 24.1500)(  2.9000,-30.3000)
% ELLIPSE 2 0 3 0
% 4 4300 1245 3500 1045 3500 1045 3500 1045
% 
\special{pn 8}%
\special{ar 4300 1246 800 200  0.0000000 6.2831853}%
% STR 2 0 3 0
% 3 4300 745 4300 845 5 0
% \LARGE $x_{0}$
\put(43.0000,-8.4500){\makebox(0,0){\LARGE $x_{0}$}}%
% ELLIPSE 2 0 3 0
% 4 2300 2650 3100 2850 730 2650 3330 2650
% 
\special{pn 8}%
\special{ar 2300 2650 800 200  6.2831853 6.2831853}%
\special{ar 2300 2650 800 200  0.0000000 3.1415927}%
% LINE 2 0 3 0
% 4 1500 1250 1500 2650 3100 1250 3100 2650
% 
\special{pn 8}%
\special{pa 1500 1250}%
\special{pa 1500 2650}%
\special{fp}%
\special{pa 3100 1250}%
\special{pa 3100 2650}%
\special{fp}%
% STR 2 0 3 0
% 3 1300 600 1300 700 5 0
% \LARGE $U$
\put(13.0000,-7.0000){\makebox(0,0){\LARGE $U$}}%
% ELLIPSE 2 0 3 0
% 4 2300 1250 1500 1050 1500 1050 1500 1050
% 
\special{pn 8}%
\special{ar 2300 1250 800 200  0.0000000 6.2831853}%
% ELLIPSE 2 0 3 0
% 4 4300 1250 5100 2850 2900 1250 5100 1250
% 
\special{pn 8}%
\special{ar 4300 1250 800 1600  6.2831853 6.2831853}%
\special{ar 4300 1250 800 1600  0.0000000 3.1415927}%
% ELLIPSE 1 0 3 0
% 4 2300 1390 2900 2590 1250 1390 2900 1390
% 
\special{pn 13}%
\special{ar 2300 1390 600 1200  6.2831853 6.2831853}%
\special{ar 2300 1390 600 1200  0.0000000 3.1415927}%
% STR 2 0 3 0
% 3 1100 2760 1100 2860 5 0
% \LARGE $U_{\rm T}$
\put(11.0000,-28.6000){\makebox(0,0){\LARGE $U_{\rm T}$}}%
% STR 2 0 3 0
% 3 1100 2490 1100 2590 5 0
% \LARGE $U_{0}$
\put(11.0000,-25.9000){\makebox(0,0){\LARGE $U_{0}$}}%
% STR 2 0 3 0
% 3 2300 750 2300 850 5 0
% \LARGE $x_{4}$
\put(23.0000,-8.5000){\makebox(0,0){\LARGE $x_{4}$}}%
% STR 2 0 3 0
% 3 2300 2100 2300 2200 5 0
% \LARGE ${\rm D8}\text{-}\overline{\rm D8}$
\put(23.0000,-22.0000){\makebox(0,0){\LARGE ${\rm D8}\text{-}\overline{\rm D8}$}}%
% VECTOR 2 0 3 0
% 2 1300 3030 1300 830
% 
\special{pn 8}%
\special{pa 1300 3030}%
\special{pa 1300 830}%
\special{fp}%
\special{sh 1}%
\special{pa 1300 830}%
\special{pa 1280 898}%
\special{pa 1300 884}%
\special{pa 1320 898}%
\special{pa 1300 830}%
\special{fp}%
% LINE 2 2 3 0
% 2 1300 2590 5500 2590
% 
\special{pn 8}%
\special{pa 1300 2590}%
\special{pa 5500 2590}%
\special{dt 0.045}%
% ELLIPSE 2 0 3 0
% 4 2300 1450 3100 1650 1500 1650 3100 1650
% 
\special{pn 8}%
\special{ar 2300 1450 800 200  0.7853982 2.3561945}%
% SARROW 2 0 3 1
% 2 1752 1596 1734 1591
% 
\special{pn 8}%
\special{pa 1752 1596}%
\special{pa 1734 1592}%
\special{fp}%
\special{sh 1}%
\special{pa 1734 1592}%
\special{pa 1794 1628}%
\special{pa 1786 1606}%
\special{pa 1804 1590}%
\special{pa 1734 1592}%
\special{fp}%
% SARROW 2 0 3 2
% 2 2848 1596 2866 1591
% 
\special{pn 8}%
\special{pa 2848 1596}%
\special{pa 2866 1592}%
\special{fp}%
\special{sh 1}%
\special{pa 2866 1592}%
\special{pa 2796 1590}%
\special{pa 2816 1606}%
\special{pa 2808 1628}%
\special{pa 2866 1592}%
\special{fp}%
% STR 2 0 3 0
% 3 2300 1700 2300 1800 5 0
% \LARGE $L$
\put(23.0000,-18.0000){\makebox(0,0){\LARGE $L$}}%
% VECTOR 2 0 3 0
% 4 2900 1250 3100 1250 2500 1250 2300 1250
% 
\special{pn 8}%
\special{pa 2900 1250}%
\special{pa 3100 1250}%
\special{fp}%
\special{sh 1}%
\special{pa 3100 1250}%
\special{pa 3034 1230}%
\special{pa 3048 1250}%
\special{pa 3034 1270}%
\special{pa 3100 1250}%
\special{fp}%
\special{pa 2500 1250}%
\special{pa 2300 1250}%
\special{fp}%
\special{sh 1}%
\special{pa 2300 1250}%
\special{pa 2368 1270}%
\special{pa 2354 1250}%
\special{pa 2368 1230}%
\special{pa 2300 1250}%
\special{fp}%
% STR 2 0 3 0
% 3 2700 1150 2700 1250 5 0
% \LARGE $R$
\put(27.0000,-12.5000){\makebox(0,0){\LARGE $R$}}%
% VECTOR 2 0 3 0
% 4 4900 1250 5100 1250 4500 1250 4300 1250
% 
\special{pn 8}%
\special{pa 4900 1250}%
\special{pa 5100 1250}%
\special{fp}%
\special{sh 1}%
\special{pa 5100 1250}%
\special{pa 5034 1230}%
\special{pa 5048 1250}%
\special{pa 5034 1270}%
\special{pa 5100 1250}%
\special{fp}%
\special{pa 4500 1250}%
\special{pa 4300 1250}%
\special{fp}%
\special{sh 1}%
\special{pa 4300 1250}%
\special{pa 4368 1270}%
\special{pa 4354 1250}%
\special{pa 4368 1230}%
\special{pa 4300 1250}%
\special{fp}%
% STR 2 0 3 0
% 3 4700 1150 4700 1250 5 0
% \LARGE $R'$
\put(47.0000,-12.5000){\makebox(0,0){\LARGE $R'$}}%
% ELLIPSE 2 0 3 0
% 4 4300 2390 5100 2590 3740 2590 4860 2590
% 
\special{pn 8}%
\special{ar 4300 2390 800 200  0.9600704 2.1815223}%
% LINE 1 2 3 0
% 60 4860 2400 3870 1410 4830 2430 3800 1400 4810 2470 3720 1380 4790 2510 3630 1350 4760 2540 3500 1280 4720 2560 3510 1350 4670 2570 3510 1410 4610 2570 3510 1470 4560 2580 3510 1530 4500 2580 3520 1600 4440 2580 3530 1670 4390 2590 3540 1740 4330 2590 3550 1810 4270 2590 3570 1890 4210 2590 3590 1970 4140 2580 3610 2050 4080 2580 3640 2140 4020 2580 3670 2230 3950 2570 3720 2340 3880 2560 3780 2460 4880 2360 3940 1420 4890 2310 4010 1430 4910 2270 4070 1430 4930 2230 4140 1440 4950 2190 4200 1440 4960 2140 4260 1440 4980 2100 4320 1440 4990 2050 4380 1440 5000 2000 4440 1440 5010 1950 4490 1430
% 
\special{pn 13}%
\special{pa 4860 2400}%
\special{pa 3870 1410}%
\special{dt 0.045}%
\special{pa 4830 2430}%
\special{pa 3800 1400}%
\special{dt 0.045}%
\special{pa 4810 2470}%
\special{pa 3720 1380}%
\special{dt 0.045}%
\special{pa 4790 2510}%
\special{pa 3630 1350}%
\special{dt 0.045}%
\special{pa 4760 2540}%
\special{pa 3500 1280}%
\special{dt 0.045}%
\special{pa 4720 2560}%
\special{pa 3510 1350}%
\special{dt 0.045}%
\special{pa 4670 2570}%
\special{pa 3510 1410}%
\special{dt 0.045}%
\special{pa 4610 2570}%
\special{pa 3510 1470}%
\special{dt 0.045}%
\special{pa 4560 2580}%
\special{pa 3510 1530}%
\special{dt 0.045}%
\special{pa 4500 2580}%
\special{pa 3520 1600}%
\special{dt 0.045}%
\special{pa 4440 2580}%
\special{pa 3530 1670}%
\special{dt 0.045}%
\special{pa 4390 2590}%
\special{pa 3540 1740}%
\special{dt 0.045}%
\special{pa 4330 2590}%
\special{pa 3550 1810}%
\special{dt 0.045}%
\special{pa 4270 2590}%
\special{pa 3570 1890}%
\special{dt 0.045}%
\special{pa 4210 2590}%
\special{pa 3590 1970}%
\special{dt 0.045}%
\special{pa 4140 2580}%
\special{pa 3610 2050}%
\special{dt 0.045}%
\special{pa 4080 2580}%
\special{pa 3640 2140}%
\special{dt 0.045}%
\special{pa 4020 2580}%
\special{pa 3670 2230}%
\special{dt 0.045}%
\special{pa 3950 2570}%
\special{pa 3720 2340}%
\special{dt 0.045}%
\special{pa 3880 2560}%
\special{pa 3780 2460}%
\special{dt 0.045}%
\special{pa 4880 2360}%
\special{pa 3940 1420}%
\special{dt 0.045}%
\special{pa 4890 2310}%
\special{pa 4010 1430}%
\special{dt 0.045}%
\special{pa 4910 2270}%
\special{pa 4070 1430}%
\special{dt 0.045}%
\special{pa 4930 2230}%
\special{pa 4140 1440}%
\special{dt 0.045}%
\special{pa 4950 2190}%
\special{pa 4200 1440}%
\special{dt 0.045}%
\special{pa 4960 2140}%
\special{pa 4260 1440}%
\special{dt 0.045}%
\special{pa 4980 2100}%
\special{pa 4320 1440}%
\special{dt 0.045}%
\special{pa 4990 2050}%
\special{pa 4380 1440}%
\special{dt 0.045}%
\special{pa 5000 2000}%
\special{pa 4440 1440}%
\special{dt 0.045}%
\special{pa 5010 1950}%
\special{pa 4490 1430}%
\special{dt 0.045}%
% LINE 1 2 3 1
% 24 5030 1910 4550 1430 5040 1860 4610 1430 5050 1810 4660 1420 5050 1750 4710 1410 5060 1700 4760 1400 5070 1650 4810 1390 5080 1600 4860 1380 5080 1540 4910 1370 5090 1490 4960 1360 5090 1430 5000 1340 5090 1370 5040 1320 5100 1320 5070 1290
% 
\special{pn 13}%
\special{pa 5030 1910}%
\special{pa 4550 1430}%
\special{dt 0.045}%
\special{pa 5040 1860}%
\special{pa 4610 1430}%
\special{dt 0.045}%
\special{pa 5050 1810}%
\special{pa 4660 1420}%
\special{dt 0.045}%
\special{pa 5050 1750}%
\special{pa 4710 1410}%
\special{dt 0.045}%
\special{pa 5060 1700}%
\special{pa 4760 1400}%
\special{dt 0.045}%
\special{pa 5070 1650}%
\special{pa 4810 1390}%
\special{dt 0.045}%
\special{pa 5080 1600}%
\special{pa 4860 1380}%
\special{dt 0.045}%
\special{pa 5080 1540}%
\special{pa 4910 1370}%
\special{dt 0.045}%
\special{pa 5090 1490}%
\special{pa 4960 1360}%
\special{dt 0.045}%
\special{pa 5090 1430}%
\special{pa 5000 1340}%
\special{dt 0.045}%
\special{pa 5090 1370}%
\special{pa 5040 1320}%
\special{dt 0.045}%
\special{pa 5100 1320}%
\special{pa 5070 1290}%
\special{dt 0.045}%
% LINE 2 2 3 0
% 2 1300 2860 5500 2860
% 
\special{pn 8}%
\special{pa 1300 2860}%
\special{pa 5500 2860}%
\special{dt 0.045}%
\end{picture}%
} \\
\end{tabular}
{\bf Fig. 2}\; The $\text{D8-}\overline{\text{D8}}$-branes configurations at high temperature. \\[10mm]
\end{center}

In the low temperature, the $\text{D8-}$ and $\overline{\text{D8}}$-branes are 
connected at $U=U_{0}$ as shown in Fig. 1. The connected 
configuration of the $\text{D8-}\overline{{\text{D8}}}$-branes indicates that 
the ${\rm U}({\rm N}_{f})_{L} \times {\rm U}({\rm N}_{f})_{R}$ global chiral symmetry is broken to a diagonal subgroup ${\rm U}({\rm N}_{f})$. We refer to the  connected configuration in the low temperature as the low-temperature phase. 
In the high temperature, there are two kinds of configurations as 
shown in Fig. 2. One is connected configuration and the other is disconnected 
configuration that the $\text{D8-}$ and $\overline{\text{D8}}$-branes hang 
vertically from infinity down to the horizon. The disconnected configuration 
of the $\text{D8-}\overline{{\text{D8}}}$-branes indicates that the 
${\rm U}({\rm N}_{f})_{L} \times {\rm U}({\rm N}_{f})_{R}$ global chiral 
symmetry is restored. We refer to the disconnected configuration and the 
connected configuration in the high temperature as the high-temperature phase 
and the intermediate-temperature phase, respectively. The intermediate-temperature phase is realized when the confinement/deconfinement phase transition and 
the chiral phase transition does not occur simultaneously.

When turning on a NS-NS $B$-field on the D-brane worldvolume, the low-energy 
effective worldvolume theories are deformed to a noncommutative Yang-Mills 
theories \cite{CDS, DH, AASJ, SJ, SW}. The D-brane realizations of 
noncommutative Yang-Mills theories have a gravity dual in the large $N$, 
strong 't Hooft coupling limit \cite{HI, MR, AOSJ}. 
In accordance with the formulation of 
\cite{HI, MR, AOSJ}, we attempt to construct the gravity dual of the noncommutative QCD whose chiral symmetry is spontaneously broken by deforming the 
holographic QCD model. Let us consider the D4-branes solution compactified on a circle in the $\tau$-direction. T-dualizing it along $x^{3}$ produces a 
D3-branes delocalized along $x^{3}$. After rotating the D3-branes along the 
($x^{2},\;x^{3}$) plane, we T-dualize back on $x^{3}$. This procedure yields 
the solution with a $B_{23}$ fields along the $x^{2}$ and $x^{3}$ directions. 
The solution in the low temperature takes the form 
\begin{align}
ds^{2} &=\left(\frac{U}{R_{\rm D4}}\right)^{3/2}
\Bigl((dt_{E})^{2}+(dx^{1})^{2}
+ h\{(dx^{2})^{2}+(dx^{3})^{2}\} + f_{\rm KK}(U)\,d\tau^{2} \Bigr) 
\nonumber \\
&+ \left(\frac{R_{\rm D4}}{U}\right)^{3/2}
\left(\frac{dU^{2}}{f_{\rm KK}(U)}+U^{2}d\Omega_{4}^{2} \right)\,,
%\nonumber \\
%& R_{\rm D4}^{3}=\pi g_{s}N_{c}l^{3}_{s}\,, \qquad 
%\tilde{f}(U)=1-\frac{U^{3}_{T}}{U^{3}}\,, \qquad 
%h(U)=\frac{1}{1+\theta^{3}U^{3}}\,,
\label{206}
\end{align}
where $h(U)=\dfrac{1}{1+\theta^{3}U^{3}}$ and $\theta$ denotes the 
noncommutativity parameter with dimension of $[{\rm length}]^{-1}$. When 
$\theta \neq 0$ this solution is dual to a gauge theory in which the 
coordinates $x^{2}$ and $x^{3}$ do not commute. It is obvious that this 
solution reduces to the solution (\ref{201}) with Euclidean signature when 
$\theta =0$. In the high temperature, the solution (\ref{206}) changes to 
\begin{align}
ds^{2} &=\left(\frac{U}{R_{\rm D4}}\right)^{3/2}
\Bigl(f_{\rm T}(U)(dx_{E})^{2}+(dx^{1})^{2}
+ h\{(dx^{2})^{2}+(dx^{3})^{2}\} + d\tau^{2} \Bigr) 
\nonumber \\
&+ \left(\frac{R_{\rm D4}}{U}\right)^{3/2}
\left(\frac{dU^{2}}{f_{\rm T}(U)}+U^{2}d\Omega_{4}^{2} \right)\,. 
\label{207}
\end{align}
The solution has the same form as the one in the low temperature (\ref{206}), but with the role of the $\tau$ and $t_{E}$ directions exchanged. 

%
%%%%%%%%%%%%%%%%%%%%%%%%%%%%%%%%%%%%%%%%%%%%%%%%%%%%%%%%%%%%%%%%%%%%%%%%%%%%%%%%
%
\section{Mass spectrum of low spin mesons in the intermediate-temperature phase} 
\setcounter{equation}{0}
\addtocounter{enumi}{1}

In order to analyze the phases of QCD at finite temperature, we needs to 
determine the shape of the D8-brane whose effective action is given by the DBI 
action and the WZ action: 
\begin{align}
\label{301}
S^{\rm D8} &=S^{\rm D8}_{\rm DBI} + S^{\rm D8}_{\rm WZ}\,, \\ 
& S^{\rm D8}_{\rm DBI}
=T_{8}\int d^{9}x e^{-\phi}{\rm Tr}\sqrt{\det (g_{MN} + B_{MN} 
+ 2\pi \alpha'F_{MN})}\,, 
\nonumber \\
& S^{\rm D8}_{\rm WZ}
=\mu_{8}\int_{\rm D8}\,C_{3} \wedge e^{(\,\widetilde{B}+2\pi l_{s}{}^{2}F\,)}\,, 
\nonumber 
\end{align}
where $T_{8}$ is the tension of the D8-brane, $\mu_{8}$ is the D8-brane charge, $\phi$ is a dilaton and $\widetilde{B}=B_{MN}dx^{M}dx^{N}$ is a 
NS-NS $B$-field: 
\begin{align}
\label{302}
&e^{2\phi} = g_{s}{}^{2}h(U)\left(\dfrac{U}{R_{\rm D4}} \right)^{3/4} \;, 
\\[3mm] 
\label{303}
&B_{MN}(U)=\left\{
\begin{array}{ll}
\theta^{3/2}\dfrac{U^{3}}{R_{\rm D4}^{3/2}}h(U) & (M=2,\;N=3) \\
0 & (\text{others})
\end{array} \right.\;.
\end{align}

We examine the spectrum of low-spin mesons in the intermediate-temperature 
phase. Let us consider the background metric (\ref{207}). This leads to an 
induced metric on D8-brane world-volume, 
\begin{align}
\label{304}
d\bar{s}^{2}_{\rm I}
&=\left( \dfrac{U}{R_{\rm D4}} \right)^{3/2}
\Bigl(f_{\rm T}(U)(dx_{E})^{2}+(dx^{1})^{2}
+ h\{(dx^{2})^{2}+(dx^{3})^{2}\} \Bigr)  \nonumber \\
&+ \left[\left( \dfrac{U}{R_{\rm D4}} \right)^{3/2}(\tau'(U))^{2} + 
\left( \dfrac{R_{\rm D4}}{U} \right)^{3/2}\dfrac{1}{f_{\rm T}(U)} \right]\,dU^{2} 
+ \left( \dfrac{R_{\rm D4}}{U} \right)^{3/2}U^{2}d\Omega^{2}_{4} \;.
\end{align}
where $\tau'(U)$ denotes the derivative of $\tau(U)$ with respect to $U$. Expanding the (abelian) DBI action with respect to $F_{MN}$, we have 
\begin{align}
\label{305}
S_{\rm D8}^{\rm DBI}
&=T_{8}\int d^{9}x e^{-\phi}
\sqrt{\det(g_{MN}+B_{MN})}\left\{1+\dfrac{1}{4}(2\pi\alpha')^{2}g^{JK}g^{PQ}
F_{JP}F_{KQ} + {\cal O}(F^4) \right\} \;.
\end{align}
We set $A_{\alpha}\;(\alpha=5,\,6,\,7,\,8,\,9)=0$ and assume that $A_{\mu}\;(\mu=0,\,1,\,2,\,3)$ and $A_{U}$ are independent of the coordinates on the $S^4$.

Under the induced metric (\ref{304}) and the dilaton (\ref{302}), the DBI action for the zeroth order in $\alpha'$ 
\begin{align}
\label{306}
S^{\rm D8(0)}_{\rm DBI}=T_{8}\int d^{9}x
e^{-\phi}\sqrt{\det(g_{MN}+B_{MN})}
\end{align}
takes the form 
\begin{align}
\label{307}
S^{\rm D8(0)}_{\rm DBI}=\dfrac{N_{f}T_{8}V_{4}}{g_{s}}
\int d^{4}x dU 
U^{4}\sqrt{f_{\rm T}(U)\left( \tau'(U) \right)^{2}
+\left(\dfrac{R_{\rm D4}}{U}\right)^{3}}\;,
\end{align}
where $N_{f}$ is the number of the flavor and $V_{4}=\dfrac{8\pi^{2}}{3}$ denotes the volume form of the unit 4-sphere. The action (\ref{307}) does not depend 
on the noncommutativity parameter $\theta$, though the action (\ref{306}) 
includes the $B$-field (\ref{303}). The equation of motion for $\tau$ is 
\begin{align}
\label{308}
\dfrac{d}{dU}\left[\dfrac{U^{4}f_{\rm T}(U)\left( \tau'(U) \right)}
{\sqrt{f_{\rm T}(U)\left(\tau'(U) \right)^{2}
+\left(\dfrac{R_{\rm D4}}{U}\right)^{3}}}\right] = 0 \;.
\end{align}
We impose the condition $\lim_{U \to U_{0}}\tau'(U) \to \infty$ that 
corresponds to the connected configuration of the $\text{D8-}
\overline{\text{D8}}$-branes.

The solution of (\ref{308}) under the condition $\lim_{U \to U_{0}}\tau'(U) \to \infty$ is given by 
\begin{align}
\label{309}
\tau(U)=U_{0}^{4}\sqrt{f_{\rm K}(U_{0})}\int^{U}_{U_{0}}
dU\,\dfrac{\left(\dfrac{R_{\rm D4}}{U}\right)^{3/2}}
{\sqrt{f_{\rm K}(U)}\sqrt{U^{8}f_{\rm K}(U)-U_{0}^{8}f_{\rm K}(U_{0})}} \;.
\end{align}
Using the solution (\ref{309}), we have 
\begin{align}
\label{310}
\left(\dfrac{U}{R_{\rm D4}}\right)^{3/2}(\tau'(U))^{2}
+ \left(\dfrac{R_{\rm D4}}{U}\right)^{3/2}\dfrac{1}{f_{\rm T}(U)}
= \left(\dfrac{R_{\rm D4}}{U}\right)^{3/2}
\dfrac{U^{8}}{U^{8}f_{\rm T}(U)-U_{0}^{8}f_{\rm T}(U_{0})}\;.
\end{align}

The DBI action (\ref{305}) with second-order in $\alpha'$ can be rewritten by 
substituting (\ref{307}) as 
\begin{align}
\label{311}
S^{\rm D8(2)}_{\rm DBI}&=
\dfrac{N_{f}T_{8}V_{4}}{g_{s}}\dfrac{(2\pi l_{s}^{2})^{2}}{2}
%\dfrac{1}{6}\dfrac{N_{f}N_{c}}{(2\pi)^{3}\alpha'R_{\rm D4}^{3}}
\int d^{4}x dU f_{\rm T}(U)^{1/2}U^{4} \left(\dfrac{R_{\rm D4}}{U}\right)^{3/2}
k(U)^{1/2} \nonumber \\
& \times \biggl[ \left( \dfrac{R_{\rm D4}}{U} \right)^{3}
f_{\rm T}(U)^{-1} \left\{ F_{01}^{2}+\dfrac{1}{h(U)}(F_{02}^{2}+F_{03}^{2}) 
\right\} \nonumber \\
& + \dfrac{1}{k(U)f_{\rm T}(U)}F_{02}^{2}+\dfrac{1}{k(U)^{2}}F_{1U}^{2}
+ \dfrac{1}{k(U)h(U)}(F_{2U}^{2}+F_{3U}^{2}) \biggr] \;.
\end{align}
where $k(U)=\dfrac{U^{8}}{U^{8}f_{\rm T}(U)-U_{0}^{8}f_{\rm T}(U)}$. 
%and $\lambda$ denotes the 't Hooft coupling. 
The WZ action is rewritten as 
\begin{align}
\label{312}
S^{{\rm D8(2)}}_{{\rm WZ}}
= \dfrac{1}{3}\dfrac{N_{c}N_{f}}{(2\pi)^{3}l_{s}^{2}}
\int d^{4}x dU B_{23} \left\{ A_{0}F_{1U} -  A_{1}F_{0U} + A_{U}F_{01} \right\}
\;.
\end{align}

The total action is given as follows 
\begin{align}
\label{313}
& S^{\rm D8(2)}=S^{\rm D8(2)}_{\rm DBI}+S^{{\rm D8(2)}}_{{\rm WZ}} \nonumber \\
&= \dfrac{N_{f}T_{8}V_{4}}{g_{s}}\dfrac{(2\pi l_{s}^{2})^{2}}{2}
%&= \dfrac{N_{c}N_{f}}{6(2\pi)^{3}\alpha'}
\int d^{4}x dU \biggl[ \left( \dfrac{R_{\rm D4}}{U} \right)^{3/2}
k(U)^{1/2}f_{\rm T}(U)^{-1/2} 
\left\{ F_{01}^{2}+\dfrac{1}{h(U)}(F_{02}^{2}+F_{03}^{2}) 
\right\} \nonumber \\
& + \left( \dfrac{U}{R_{\rm D4}} \right)^{3}
k(U)^{-1/2}\left\{f_{\rm T}(U)^{-1/2}F_{0U}^{2}+f_{\rm T}(U)^{1/2}F_{1U}^{2}
+\dfrac{f_{\rm T}(U)^{1/2}}{h(U)}(F_{2U}^{2}+F_{3U}^{2}) \right\} \nonumber \\
& + \kappa \left( \dfrac{U}{R_{\rm D4}} \right)^{3/2}
\dfrac{N(\Theta, U)}{(U/U_{0})^{5/2}}\left\{A_{0}F_{1U}-A_{1}F_{0U}+A_{U}F_{01}
\right\}
\biggr] \;,
\end{align}
where $\kappa = 2/U_{0}$, $N(\Theta, U)=\dfrac{\Theta^{3/2}(U/U_{0})^{3}}{1+\Theta^{3}(U/U_{0})^{3}}$ and $\Theta=U_{0}\theta$ denotes the dimensionless 
noncommutativity parameter. In deriving the coefficient of (\ref{313}), we 
utilize $T_{8}=\dfrac{1}{(2\pi)^{8}l_{s}^{9}}$ and $R_{\rm D4}^{3}=\pi d_{s}N_{c}l_{s}^{3}$. 

To simplify the consideration, we choose to focus on the components $A_{1}$ and $A_{U}$, and suppose that these gauge fields do not depend on the spatial 
coordinates $x_{1},\;x_{2}$ and $x_{3}$. The equations of motion for $A_{1}$ 
and $A_{U}$ are given by  
\begin{align}
\label{314}
& \partial_{U} 
\left\{2k(U)^{-1/2}f_{\rm T}(U)^{1/2}\left( \dfrac{U}{R_{\rm D4}} \right)^{3/2}
\partial_{U}A_{1} \right\} 
- 2k(U)^{1/2}f_{\rm T}(U)^{-1/2}\left( \dfrac{R_{\rm D4}}{U} \right)^{3/2}
\partial_{0}^{2}A_{1} \nonumber \\ 
& + 2\kappa \left( \dfrac{U}{R_{\rm D4}} \right)^{3/2} 
\dfrac{N(\Theta, U)}{(U/U_{0})^{5/2}}F_{0U}
-\partial_{U}\left\{\kappa \left( \dfrac{U}{R_{\rm D4}} \right)^{3/2}
\dfrac{N(\Theta, U)}{(U/U_{0})^{5/2}} \right\}A_{0}=0 \;,\\
\label{315}
& f_{\rm T}(U)^{-1/2}\partial_{0}F_{0U} 
- \kappa k(U)^{1/2} (U/U_{0})^{-5/2}\dfrac{N(\Theta, U)}{(U/U_{0})^{5/2}}
\partial_{0}A_{1}=0 \;.
\end{align}
Eliminating $F_{0U}$ from (\ref{314}) and (\ref{315}), we have 
\begin{align}
\label{316}
& \partial_{U} 
\left\{2k(U)^{-1/2}f_{\rm T}(U)^{1/2}U^{4}
\left( \dfrac{R_{\rm D4}}{U} \right)^{3/2}
\partial_{U}\varphi \right\} - 2k(U)^{1/2}f_{\rm T}(U)^{-1/2}U^{4}
\left( \dfrac{R_{\rm D4}}{U} \right)^{9/2}
\partial_{0}^{2}\varphi \nonumber \\ 
&+ 2\kappa^{2}k(U)^{1/2}f_{\rm T}(U)^{1/2}U^{4}
\left( \dfrac{R_{\rm D4}}{U} \right)^{3/2}
\dfrac{N(\Theta, U)^{2}}{(U/U_{0})^{5}}\varphi=0 \;.
\end{align}
In deriving the equation of motion, we have put $\varphi=\partial_{0}A_{1}$ and taken the Coulomb gauge $A_{0} \approx 0$. We expand the field $\varphi$ in terms of the complete set $\{\psi_{n}(U)\}\;\;(n=1,\,2,\,3,\,\ldots)$ as $\varphi(t,\;U)=\sum_{n}v^{(n)}(t)\psi_{n}(U)$. The masses of $v^{(n)}$ are defined by
\begin{align}
\label{317}
\partial_{0}^{2}v^{(n)}+m_{n}^{2}v^{(n)}=0 \;.
\end{align}
Using the equations (\ref{317}), we derive the equation for the modes $\psi_{n}$%
\begin{align}
\label{318}
& u^{1/2}k(u)^{-1/2}f_{\rm T}(u)^{1/2}\partial_{u}
\left\{ u^{5/2} k(u)^{-1/2} f_{\rm T}(u)^{1/2} \partial_{u}\psi_{n} \right\} 
\nonumber \\
& + \left[\dfrac{R_{\rm D4}^{3}}{U_{0}}m_{n}^{2} + 4u^{-2}f_{\rm T}(u)
N(\Theta, u)^{2} \right]=0 \;,
\end{align}
where we have used the dimensionless variables $u=U/U_{0}$. The functions $f_{\rm T}(u),\;k(u)$ and $N(\Theta, u)$ are rewritten with respect to the dimensionless variables $u$ as follows; 
\begin{align}
\label{319}
& f_{\rm T}(u)=1-\left( \dfrac{u_{\rm T}}{u} \right)^{3} \;, \qquad 
k(u)=\dfrac{u^{8}}{u^{8}f_{\rm T}(u)-f_{\rm T}(1)}\;, 
\nonumber \\
& N(\Theta, u)= \dfrac{\Theta^{3/2}u^{3}}{1+\Theta^{3}u^{3}}\;,
\end{align}
where $u_{\rm T}=U_{\rm T}/U_{0}$ is a dimensionless parameter. 

Eq. (\ref{319}) is similar to the equation at finite temperature case \cite{PZ}. The difference is the appearance of extra term $N(\Theta, u)$ in the mass term. It is convenient to decompose Eq. (\ref{319}) into the first-order differential equations as 
\begin{align}
\label{320}
\left\{ 
\begin{array}{ll}
\partial_{z}\psi_{n}(z)
&=\dfrac{2}{3}\dfrac{z(1+z^2)^{1/3}}{
\{(1+z^2)^{8/3}-1 -u_{\rm T}{}^{3}((1+z^2)^{5/3}-1)\}^{1/2}
(1+z^2-u_{\rm T}^{3})^{1/2}} \,\zeta_{n}(z) \\[5mm] 
\partial_{z}\zeta_{n}(z)
&=-\dfrac{2}{3}\dfrac{z(1+z^2)}{
\{(1+z^2)^{8/3}-1 -u_{\rm T}{}^{3}((1+z^2)^{5/3}-1)\}^{1/2}
(1+z^2-u_{\rm T}^{3})^{1/2}} \\[3mm] 
& \hspace*{15mm} \times 
\left[\lambda_{n} + \dfrac{4\Theta^{3}(1+z^2-u_{\rm T}^3)(1+z^2)^{1/3}}
{(1+\Theta^{3}(1+z^2))^{2}}\right]
\psi_{n}(z) 
\end{array} \right.
\end{align}
where $\lambda_{n}=\dfrac{R_{\rm D4}^{3}}{U_{0}}m_{n}^{2}$ and $\zeta_{n}(z)=\dfrac{3}{2}\dfrac{\{(1+z^2)^{8/3}-1 -u_{\rm T}{}^{3}((1+z^2)^{5/3}-1)\}^{1/2}}{z(1+z^2)^{1/3}}\partial_{z}\psi_{n}(z)$. In deriving these equations, we have used the new variable $z^{2}=u^{3}-1$. As Eqs. (\ref{320}) are symmetric for the transform $z \to -z$, $\psi_{n}(z)$ is even or odd function in z-variable. We can obtain the value of dimensionless parameter $\lambda_{n}$ from Eq. (\ref{320}) under the regularity condition : $\psi_{n}(0)=0 \;\; {\rm or}\;\; \partial_{z}\psi_{n}(0)=0$. 

The asymptotic distance $L=\int d\tau$ between the position of D8 and anti-D8-branes is rewritten by using the expression (\ref{309}) as
\begin{align}
\label{321}
L=\left(\dfrac{R_{\rm D4}{}^{3}}{U_{0}}\right)^{1/2}\,K(u_{\rm T})\,,
\qquad 
K(u_{\rm T}) \equiv 2\int^{\infty}_{1}
du \dfrac{u^{-3/2}}{\sqrt{f_{\rm T}(u)}
\sqrt{\dfrac{f_{\rm T}(u)}{f_{\rm T}(1)}u^8-1}}\;.
\end{align}
Recall that the temperature $T$ is expressed in terms of the parameters $u_{\rm T}$ and $R_{\rm D4}$:
\begin{align}
\label{322}
T = \dfrac{3}{4\pi}\left(\dfrac{U_{\rm T}}{R_{\rm D4}{}^{3}}\right)^{1/2}
= \dfrac{3}{4\pi}\sqrt{u_{\rm T}}
\left(\dfrac{U_{0}}{R_{\rm D4}{}^{3}}\right)^{1/2}\;,
\end{align}
where $U_{0}$ is the minimal point of the connected probe brane configuration, 
which can be related to the asymptotic distance $L$. From (\ref{321}) and (\ref{322}), we can rewrite the temperature $T$ in terms of the asymptotic distance $L$: 
\begin{align}
\label{323}
T(u_{\rm T}) = \dfrac{3}{4\pi}\sqrt{u_{\rm T}}\,\dfrac{K(u_{\rm T})}{L}\;.
\end{align}
The ratio of $T$ to the confinement/deconfinement critical temperature $T_{c}=\dfrac{1}{2\pi R}$ is 
\begin{align}
\label{324}
\dfrac{T(u_{\rm T})}{T_{c}} = \dfrac{3}{2}\sqrt{u_{\rm T}}\,K(u_{\rm T})\,
\dfrac{R}{L}\;.
\end{align}
Substituting the $\chi$SB/SB critical temperature $T(u_{\rm T}=0.73572)$ 
into (\ref{324}), we have
\begin{align}
\label{325}
\dfrac{T(u_{\rm T}=0.73572)}{T_{\rm c}}
\cong 0.97 \,\dfrac{R}{L}\;.
\end{align}
When the ratio (\ref{324}) is larger than 1, that is $L/R<0.97$, there is an 
intermediate-temperature phase \cite{ASY}. When we put $L/R=0.4$, the 
intermediate-temperature phase occurs in the temperature range 
$1<T(u_{\rm T})/T_{\rm c}<2.42$. 

The mass squared of the mesons are expressed in terms of the eigenvalues $\lambda_{n}$ in (\ref{320}) by using (\ref{321}): 
\begin{align}
\label{326}
m_{n}{}^{2}=\dfrac{U_{0}}{R_{\rm D4}{}^{3}}\,\lambda_{n}
= \dfrac{K(u_{\rm T})^2}{L^2}\,\lambda_{n}\;.
\end{align}

\begin{center}
\begin{tabular}{cc}
%\multicolumn{2}{c}
%{\scalebox{0.8}{\includegraphics[width=100mm]{CUT(q=0.0).eps}}} \\
%\multicolumn{2}{c}{$q=0.0$} \\
\scalebox{0.8}{\includegraphics[width=100mm]{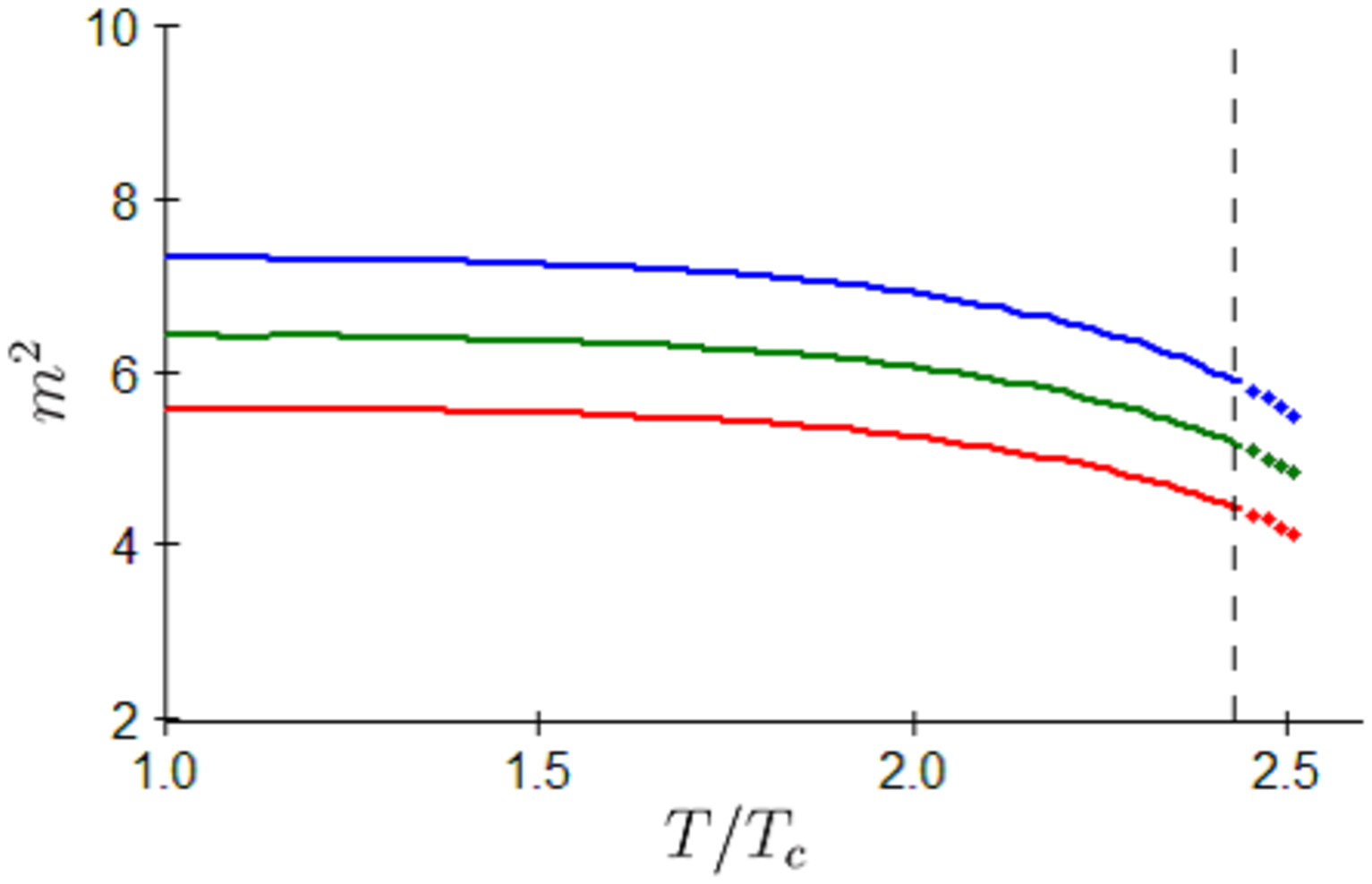}} & 
\scalebox{0.8}{\includegraphics[width=100mm]{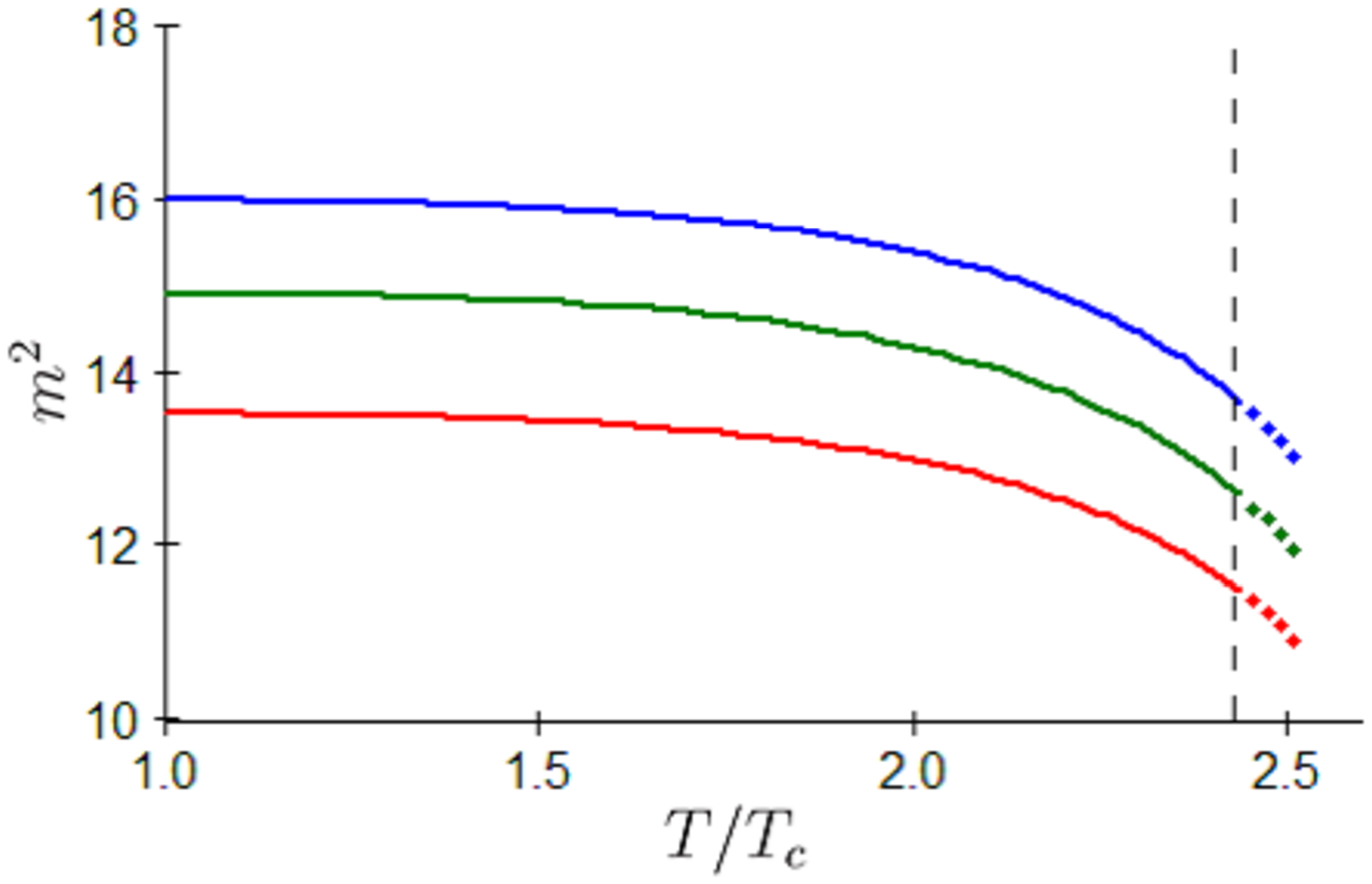}} \\
$\rho$-meson & $a_{1}$-meson 
\end{tabular}
\end{center}
{\bf Fig. 3}\; Masses squared of the lightest $\rho$(left)- and $a_{1}$(right)- mesons as a function of temperature $T$ in the intermediate-temperature regime. From the top(blue) to the bottom(red) line in the left hand side graph 
represents the value of $\Theta$ for 0.0, 1.5, 0.1, respectively. From the 
top(blue) to the bottom(red) line in the right hand side graph represents the 
value of $\Theta$ for 0.0, 1.2, 0.1, respectively. \\[8mm]

The mass squared of the mesons depends on the temperature $T$. The temperature dependence of mass squared for the lightest $\rho$(vector)- and $a_{1}$(pseudovector)- mesons  in the intermediate-temperature regime are shown in figure 3.

We can observe that the masses of light vector- and pseudovector-meson 
decrease with increasing temperature $T$. This is consistent with the result 
of the holographic model for mesons \cite{KG_MY, PSZ} and lattice calculation 
\cite{FK}. In addition, the masses of meson reduce over the whole area in the 
intermediate-temperature regime by the space noncommutativity. In this sense, 
the noncommutativity parameter $\theta$ plays the role like the temperature 
$T$. However, the dependence for the noncommutativity parameter of the masses 
is not monotonic. The noncommutativity parameter dependence of the lightest 
$\rho$\,(vector)- and $a_{1}$\,(pseudovector)- mesons  in the intermediate-
temperature regime are shown in figure 4.

\begin{center}
\begin{tabular}{cc}
\scalebox{0.8}{\includegraphics[width=100mm]{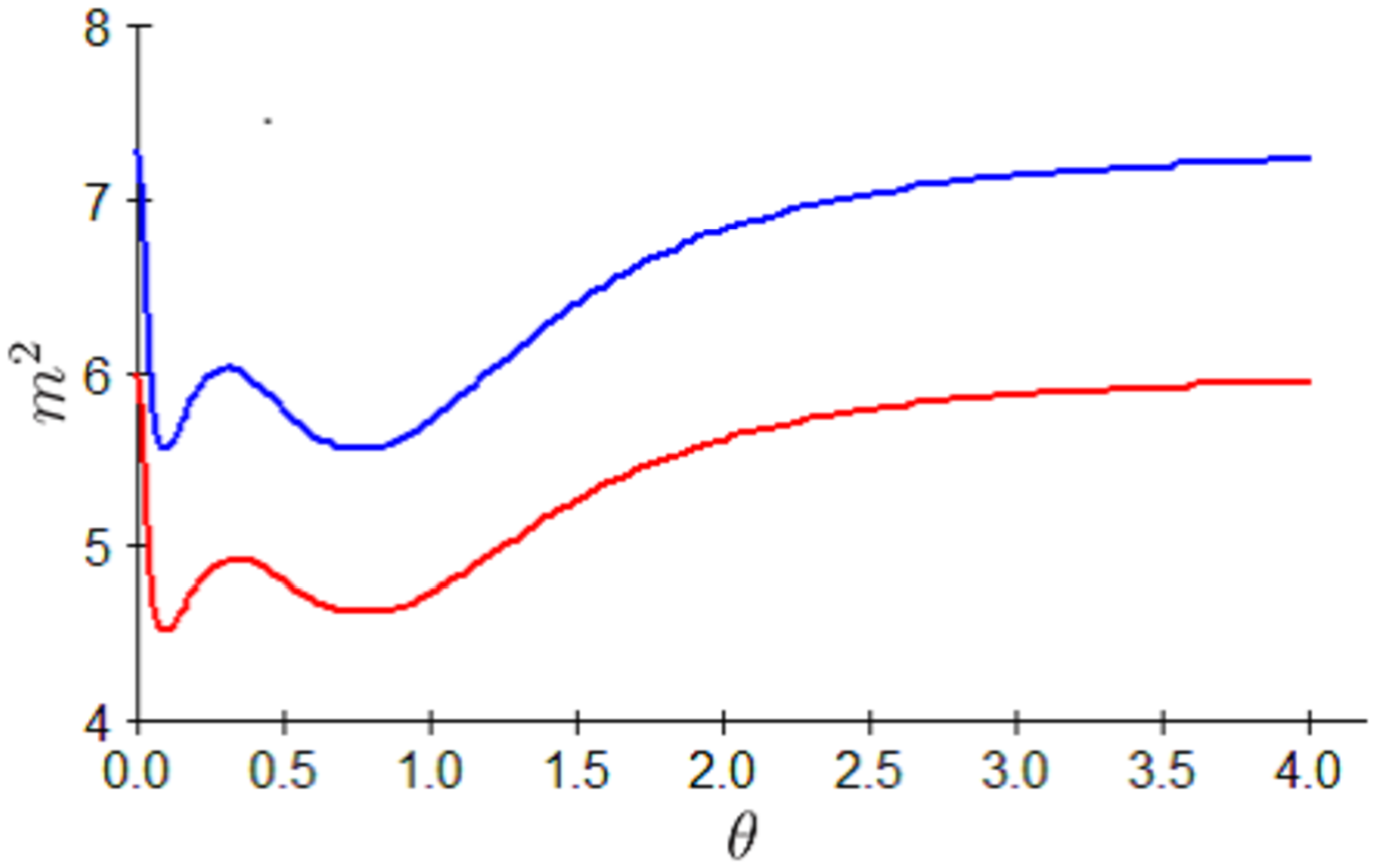}} & 
\scalebox{0.8}{\includegraphics[width=100mm]{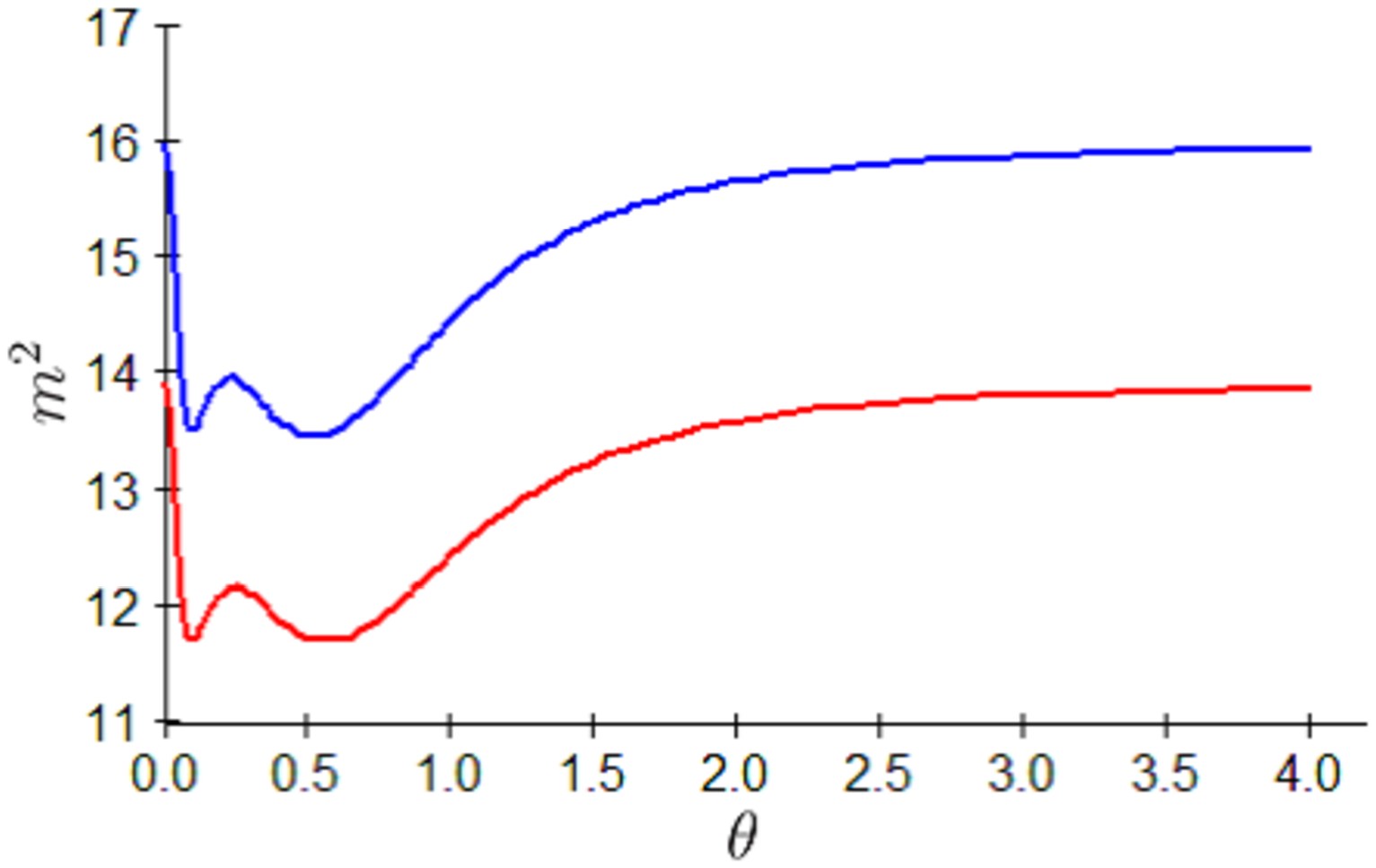}} \\
$\rho$-meson & $a_{1}$-meson 
\end{tabular}
\end{center}
{\bf Fig. 4}\; Masses squared of the lightest $\rho$(left)- and $a_{1}$(right)- mesons as a function of noncommutativity parameter $\Theta$ in the 
intermediate-temperature regime. The lower(red) and upper(blue) line in the 
each side graph represents the value of $T$ for 1.2, 2.4, respectively. 
\\[8mm]

There is a little difference of the noncommutativity parameter dependence 
between the masses of $\rho-$ and $a_{1}$ mesons. However, there is a slight 
difference of the noncommutativity parameter dependence between the masses of 
high and low temperature. We notice that the masses of light vector- and 
pseudovector-meson return to the commutative case as the noncommutativity parameter becomes larger. This feature is common with the gravity dual of 
noncommutative gauge theories \cite{APR, NST, TN_YO_KS}. The mass trajectories 
for the ground and excited state in the intermediate-temperature regime are 
shown in figure 5.

\begin{center}
\begin{tabular}{cc}
\scalebox{0.8}{\includegraphics[width=100mm]{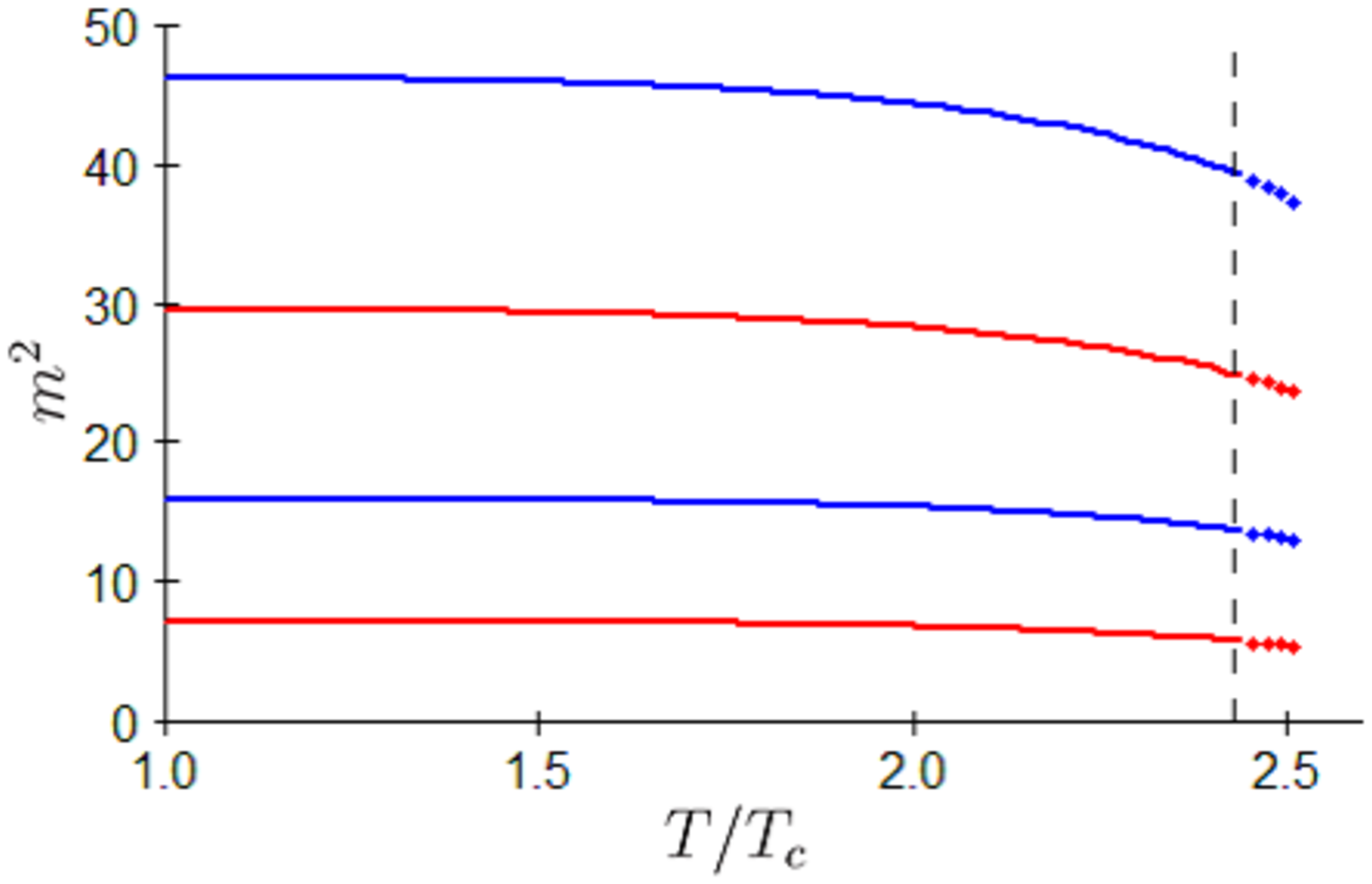}} & 
\scalebox{0.8}{\includegraphics[width=100mm]{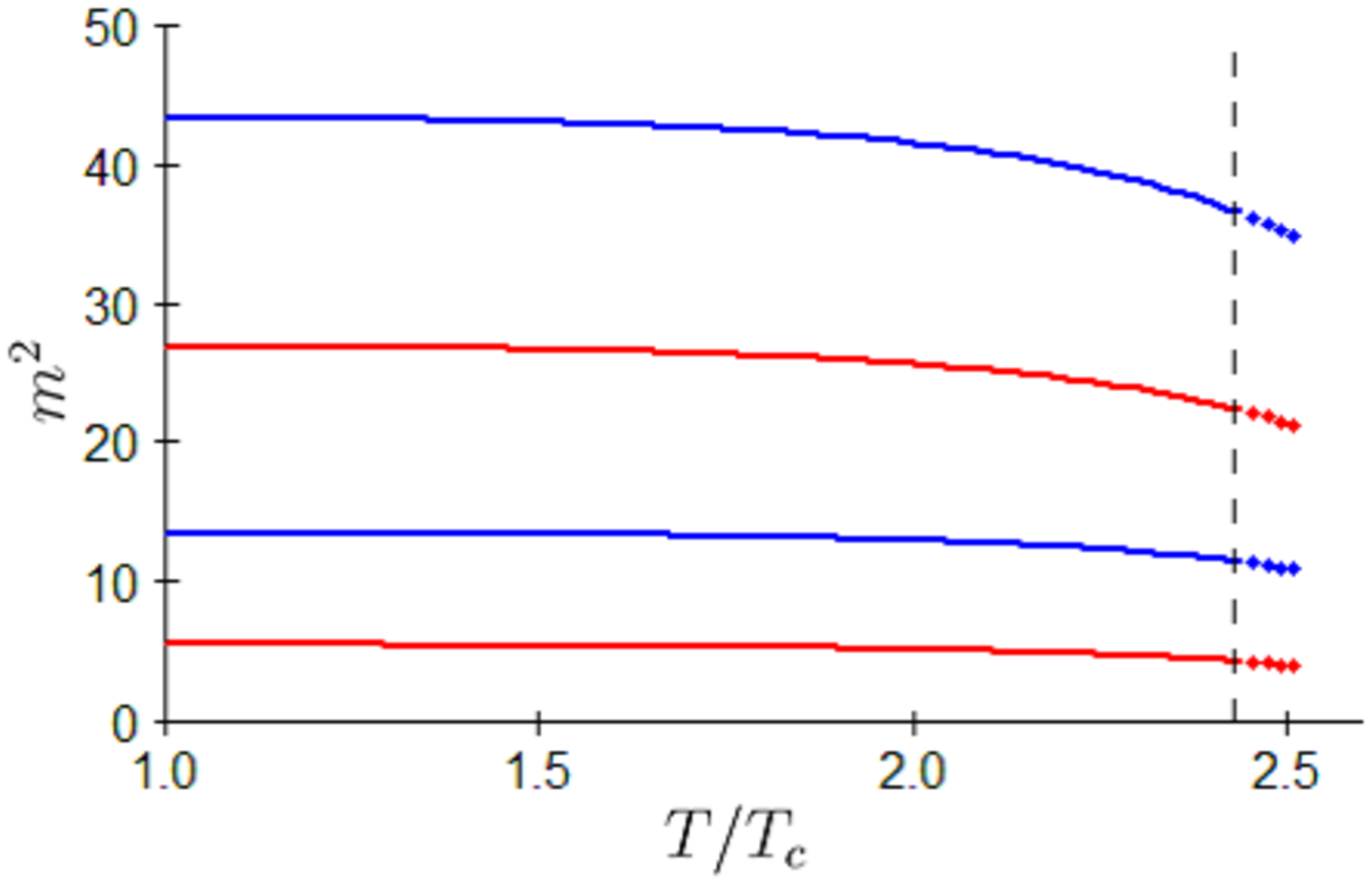}} \\
$\theta=0.0$ & $\theta=0.1$ 
\end{tabular}
\end{center}
{\bf Fig. 5}\; Masses squared of the ground and first excited state of mesons as a function of temperature $T$ in the intermediate-temperature regime. The 
commutative ($\theta=0.0$) and noncommutative ($\theta=0.0$) case is 
illustrated on the left and right hand side of the figure, respectively. From 
the bottom to the top line in the both sides graph represents the ground state 
of $\rho$-meson, the ground state of $a_{1}$-meson, the first excited state of 
$\rho$-meson, the first excited state of $a_{1}$-meson, respectively. \\[8mm]
In noncommutative case, the masses of meson decrease as temperature increase 
over the whole area in the intermediate-temperature regime not only the ground 
state but also the excited states. However, the mass differences of the 
excited state of meson between the commutative case and the noncommutative 
case in the same temperature are larger than that of the ground state. The 
mass trajectories for the ground and excited state in the noncommutative case 
become flat than that in the commutative case. 

%
%%%%%%%%%%%%%%%%%%%%%%%%%%%%%%%%%%%%%%%%%%%%%%%%%%%%%%%%%%%%%%%%%%%%%%%%%%%%%%%
%
\section{Conclusions}
\setcounter{section}{4}
\setcounter{equation}{0}
\addtocounter{enumi}{1}

In this paper, we have constructed a noncommutative deformation of the holographic QCD (Sakai-Sugimoto) model after the prescription of Alishahiha et al \cite{AOSJ} and Arean et al. \cite{APR} and evaluated the mass spectrum of low spin 
vector mesons at finite temperature. The masses of light vector- and 
pseudovector-meson in the noncommutative holographic QCD model reduces over 
the whole area in the intermediate-temperature regime compared to the 
commutative case. However, the space noncommutativity does not change the 
properties of the temperature dependence for the mass spectrum of low spin 
mesons \cite{KG_MY, PSZ}. The masses of meson also decrease with increasing 
temperature in noncommutative case. The variation of masses of meson with 
temperature becomes small in noncommutative case. The intermediate-temperature 
phase is easy to be realized in the noncommutative deformation of the 
holographic QCD model \cite{TN_YO_KS}. The smallness of mass variation with 
temperature, namely, the hardness of melting meson may be related with the 
property of phase in the noncommutative deformation of the holographic QCD 
model.

The magnitude of noncommutativity parameter $\theta$ denotes the degree of 
space noncommutativity in the noncommutative theory. The noncommutative 
deformation of the holographic QCD model reduces to the commutative one when 
the noncommutativity parameter tends to zero. However, the mass spectrum of 
low spin mesons in the noncommutative theory result in that of the commutative 
theory under the large $theta$ limit. This fact shows that the noncommutative 
deformation of the holographic QCD model also reduces to the commutative model 
when the noncommutativity parameter approaches infinity. This property is 
common to other noncommutative deformation of holographic models. In addition, 
it is also common with other noncommutative deformation of holographic models 
that noncommutativity is introduced via the WZ terms in the effective action 
for probe brane \cite{APR, TN_YO_KS}.

It has been shown that the dissociation temperature of mesons with large spin 
is spin dependent and higher spin mesons have a tendency to melt at lower 
temperature in the context of the holographic QCD model \cite{PSZ, PSZ2}. To 
investigate how the noncommutativity have an effect on the properties of meson 
melting is of particular interest. We can confirm that the dissociation 
temperature of large spin mesons also depend on the noncommutativity parameter \cite{TN_YO_KS2}.

The UV/IR mixing is well known as distinctive features of noncommutative field 
theories\cite{MRS, DN, S}. The UV/IR mixing appears to be the qualitative difference between ordinary and noncommutative field theory. The difference in 
properties of meson mass spectrum between ordinary and noncommutative 
deformation of QCD might be related to the UV/IR mixing. We hope to discuss 
this subject in the future.

\section*{Acknowledgments}

Y.O. was supported by MEXT grant to aid women researchers. T.N. would like to 
thank members of the Physics Department at College of Engineering, Nihon 
University for their encouragements. The authors are grateful to A. Sugamoto 
for useful discussions and comments. 

\clearpage 

%
%%%%%%%%%%%%%%%%%%%%%%%%%%%%%%%%%%%%%%%%%%%%%%%%%%%%%%%%%%%%%%%%%%%%%%%%%%%%%%%
%

\end{document}